\documentclass{article}

\PassOptionsToPackage{numbers,compress}{natbib}
\usepackage[preprint]{neurips_2026}

\usepackage[utf8]{inputenc}
\usepackage[T1]{fontenc}
\usepackage{hyperref}
\usepackage{url}
\usepackage{booktabs}
\usepackage{graphicx}
\usepackage{xcolor}
\usepackage{amsmath}
\usepackage{amssymb}
\usepackage{amsfonts}
\usepackage{algorithm}
\usepackage{algpseudocode}
\usepackage{array}
\usepackage{siunitx}
\usepackage{caption}
\usepackage{subcaption}
\usepackage{multirow}
\usepackage{microtype}

\newcolumntype{L}[1]{>{\raggedright\arraybackslash}p{#1}}
\graphicspath{{./figures/}}

\title{Irminsul: MLA-Native Position-Independent Caching\\
for Agentic LLM Serving}

\author{%
  Bole Ma \\
  Erlangen National High \\
  Performance Computing Center \\
  Erlangen, Germany \\
  \texttt{bole.ma@fau.de}
  \And
  Jan Eitzinger \\
  Erlangen National High \\
  Performance Computing Center \\
  Erlangen, Germany \\
  \texttt{jan.eitzinger@fau.de}
  \And
  Harald Köstler \\
  Erlangen National High \\
  Performance Computing Center \\
  Erlangen, Germany \\
  \texttt{harald.koestler@fau.de}
}

\begin{document}

\maketitle

\begin{abstract}
Agentic LLM workloads put bit-identical tokens at shifted positions every turn, voiding prefix caches at the first byte of divergence. Operators report cache-hit regressions ranging from moderate slowdowns to severe TTFT spikes of 10–16s on unchanged content~\cite{openclaw:issue40256,openclaw:issue66389}.
Prior position-independent caching systems correct RoPE on the full $d_K$-dimensional key, an architectural cost imposed by GQA, not by caching itself. Multi-Head Latent Attention, deployed at scale in
DeepSeek-V2/V3/R1~\cite{deepseekv2,deepseekv3,deepseekr1},
Kimi-K2/Moonlight~\cite{kimik2, moonlight}, GLM-5~\cite{glm5}, and
Mistral Large~3~\cite{mistrallarge3}, factors each KV row into a position-free $c_{KV}$ and a 64-dim $k_r$ correctable in closed form; this structure motivates content-addressed caching as a natural fit rather than a GQA workaround.
We present Irminsul, which extends SGLang's radix cache with content-hash keying over CDC-chunked segments and a $\delta$-rotation rule for $k_r$. We evaluate three native MLA-MoE deployments --- DeepSeek-V2-Lite~\cite{deepseekv2} ($16$B/$2.4$B), Kimi Moonlight-16B-A3B~\cite{moonlight}, and JoyAI-Flash~\cite{joyaillm} ($48$B/$3$B) --- with output-consistency on all three and recovery measured on the two endpoints; Irminsul recovers up to ${\sim}83\%$
of prompt tokens above exact-prefix on agentic traffic while delivering $63\%$ prefill energy savings per cache hit. We argue that content-addressed caching belongs in the serving stack as a first-class primitive, not a retrofit over prefix matching.
\end{abstract}

\begin{figure*}[h!]\centering
\includegraphics[width=0.95\textwidth]{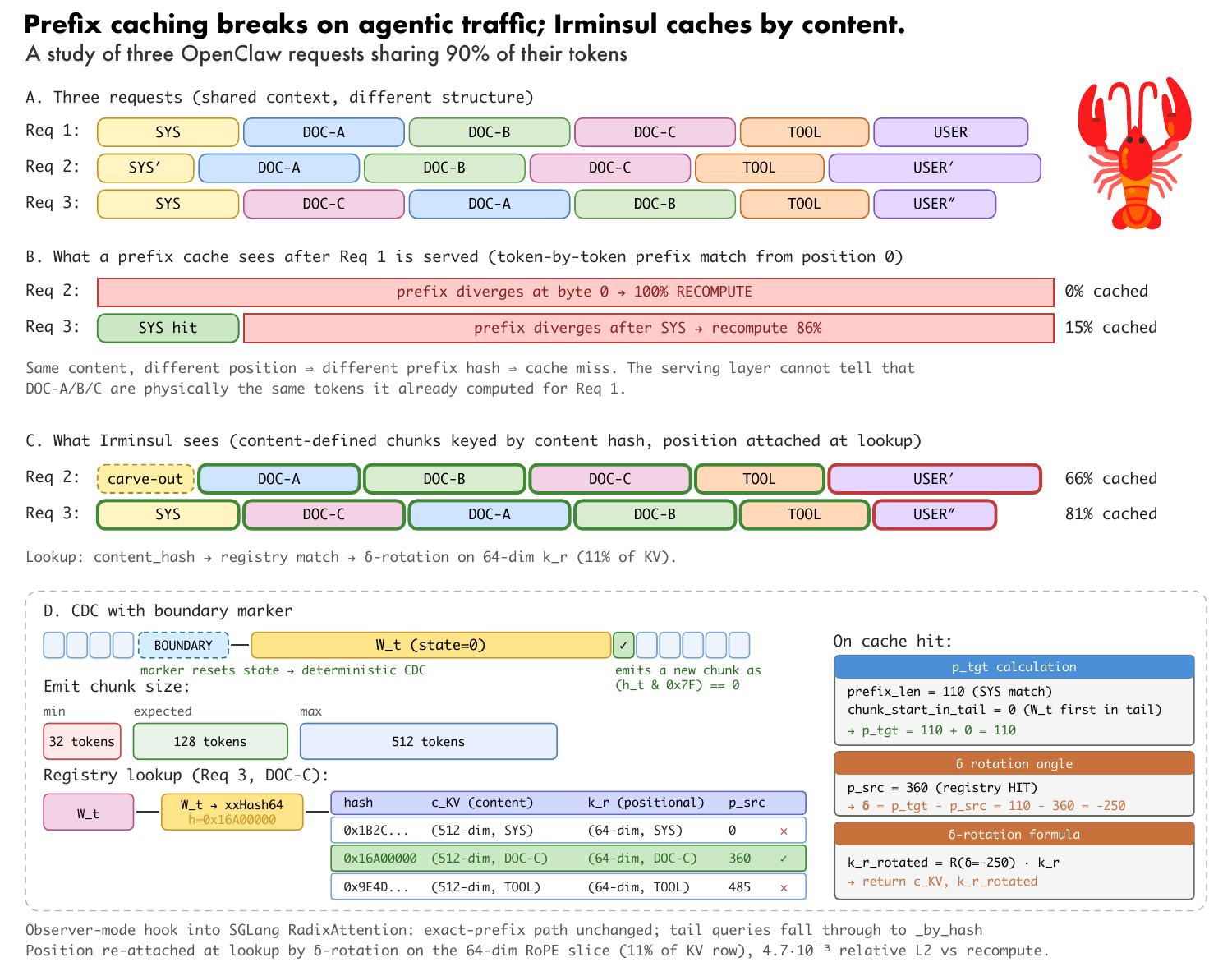}
\caption{\textbf{The problem and our fix, by example.}
  (A)~Three realistic OpenClaw-mode requests share $\sim\!90\%$ of
  their tokens but differ in ordering or in one system-variable.
  (B)~Exact-prefix cache voids every downstream block on a
  single-byte divergence: Req\,2 is $0\%$ cached, Req\,3 is $15\%$.
  (C)~Irminsul keys chunks by content hash and re-attaches position
  via $\delta$-rotation on MLA's 64-dim RoPE slice; the same requests
  become $66\%$ and $81\%$ cached with the first CDC chunk carved out.
  (D)~The rolling-hash boundary rule that produces the chunk partition.
  Measured on the paper's partition-shift A/B (Table~\ref{tab:ab}),
  Irminsul recovers $77$ percentage points above exact-prefix on
  agent-metadata rotation.}
\label{fig:teaser}
\end{figure*}

\section{Introduction}
\label{sec:intro}

\emph{Cache what you mean, not where you put it.}
Prefix caching~\cite{sglang,vllm} is the dominant prefill optimisation
in production LLM serving: when turn~$T{+}1$'s context is turn~$T$
plus one new message, KV is hashed by token sequence and served
instantly.  Agentic systems break this assumption every turn.

\textbf{The core issue: position shift is structural, not accidental.}
Every agentic turn assembles a prompt by composing a large static
system prompt, growing history, and dynamically-placed retrieved
documents and tool outputs.  The composition \emph{shifts} previously
seen content to new positions even when the tokens themselves are
unchanged.  Public OpenClaw operator post-mortems document the
consequences: retrieval re-ranking shifts document
positions~\cite{openclaw:issue40256}; a process-spawning regression
strands 40--45\,K-token history outside the cache
($100\%{\to}35\%$)~\cite{openclaw:issue66389}; forward-iterating
compaction busts cache past the $75\%$
threshold~\cite{openclaw:pr58036}.  The pattern is not specific to
OpenClaw: \emph{any} framework that composes dynamic content
(LangGraph-style orchestration, tool-use with retrieval, coding agents
with compaction) exhibits it.  Our workload analysis
(\S\ref{sec:workload}) confirms that up to 48\% of session tokens
(p95 within-session, Toolathlon) repeat at shifted positions on real agentic traffic.  The right fix is
at the serving layer, not inside any single agent.

\textbf{Why prior PIC systems cannot fully solve it.}
\emph{Position-Independent Caching} (PIC) is the principled response;
prior systems (CacheBlend, EPIC, MEPIC, KV-Packet, SemShareKV;
\S\ref{sec:background}) recover chunk-level KV across positional shifts
on GQA, but GQA fuses position inseparably into the full $K$ tensor
(algebraically correctable via an inverse rotation, but only at full-$d_K$ width per token),
leaving a residual per-hit correction cost that is architectural, not
an engineering shortcoming.

\textbf{The structural opportunity: MLA doesn't have this problem.}
Multi-Head Latent Attention~\cite{deepseekv2} factors each token's KV
into a position-free latent $c_{KV}$ and a small RoPE-carrying key
$k_r$ (structure detailed in \S\ref{sec:background}).  The bulk of the
KV row is therefore reusable verbatim at any new absolute position;
only $k_r$ needs correction, and it corrects in closed form via a
single $\delta$-rotation.  This is not merely an analytical saving but
the property that makes MLA-native PIC deployable at production latency
budgets.

\textbf{Irminsul} is our realisation of this opportunity: content-hash
keying over CDC-chunked segments, a $\delta$-rotation rule for the
RoPE slice fused into FlashMLA~\cite{flashmla}, and a first-chunk
carve-out that sidesteps the attention-sink regime
(\S\ref{sec:algo}).

\textbf{Contributions.}
(i)~\textbf{Irminsul}, the first MLA-native PIC system, with
a precision study bounding $\delta$-rotation error at
$4.7{\times}10^{-3}$ rel-L2 in bf16 and an output-consistency study
across four MLA configurations --- three native MLA-MoE deployments
at $16$B/$2.4$B, $16$B/$3$B, and $48$B/$3$B (the load-bearing scale
axis, spanning DSv2-form and DSv3-form rotaries with
$\theta\in\{10^4, 5{\times}10^4, 3.2{\times}10^7\}$) plus a 4B
retrofit stress-test (\S\ref{sec:precision});
(ii)~a $\tau$-free \textbf{ROC methodology} for PIC feasibility
(\S\ref{sec:roc}), showing MLA's NoPE component scores AUC~0.77 while
raw $K$ in pure-softmax architectures scores \emph{below random} ---
the structural argument for MLA-native PIC; and
(iii)~a cross-architecture \textbf{energy study} establishing PIC's
scope: softmax-attention caches save $63$--$86\%$ of prefill energy per
hit while three hybrid SSM families save $\approx\!0\%$
(\S\ref{sec:energy}).
We demonstrate the following supporting evidence: 1\,035-trajectory agentic workload subset
($5.1{\times}10^7$ tokens; \S\ref{sec:workload}); a recoverability
study showing CDC+fallback finds $3.8$--$4.8{\times}$ more reuse than
fixed-block hashing offline, motivating the runtime's marker-pinned
single-pass CDC (\S\ref{sec:algorithm}, \S\ref{sec:algo}); and a
measurement showing the attention sink is sequence-start-local (not
chunk-start-local), justifying the first-chunk carve-out
(\S\ref{sec:algo}).

\section{Background and Related Work}
\label{sec:background}

\textbf{Exact-prefix caching.}
SGLang's RadixAttention~\cite{sglang} and vLLM's
PagedAttention~\cite{vllm} index KV by token-sequence hash at fixed
block granularity and hit on exact prefix match --- which fits
conversation replay but misses any position-shifted reuse.

\textbf{Prior PIC.}
CacheBlend~\cite{cacheblend}, EPIC~\cite{epic}, MEPIC~\cite{mepic},
KV-Packet~\cite{kvpacket}, and SemShareKV~\cite{semsharekv} each
recover chunk-level KV across positional shifts in GQA.
Without correction the cached $K$ carries the stored position's rotation,
injecting correlated errors worse than random~\cite{cacheblend}.
Each system mitigates this: CacheBlend recomputes ${\sim}15\%$ of tokens
($O(0.15\,N^2)$); EPIC recomputes $k{\approx}32$ boundary tokens
($O(kN)$); MEPIC eliminates per-hit recompute via a bespoke kernel, but
one that is GQA-specific and does not transfer to other KV structures.
None has been evaluated on MLA.

\textbf{MLA.}
MLA~\cite{deepseekv2,deepseekv3,deepseekr1,kimik2,glm5,mistrallarge3,moonlight} stores per-token
a position-free $c_{KV}$ latent ($512$ dims) and a decoupled
RoPE key $k_r$ ($64$ dims); full $K=[W_{uk}\,c_{KV},\,k_r]$ and $V$
is fully derived from $c_{KV}$.  SnapMLA~\cite{snapmla} notes the
two-component structure but not the caching implication; vLLM V1
treats an MLA KV block as opaque.

\textbf{Hybrid linear/SSM.}
Mamba2~\cite{mamba2}, GDN~\cite{gdn}, and KDA~\cite{kdalinear}
replace quadratic attention with a bounded recurrent state, whose
monolithic structure makes PIC structurally inapplicable at the
recurrent layers~\cite{marconi}; \S\ref{sec:energy} confirms the
zero-savings consequence quantitatively.

\section{Agentic Workload Characterisation}
\label{sec:workload}

\emph{Before building a system to exploit position-shifted reuse, we
need to establish how much of it exists in real agentic traffic.}
If the reuse fraction is small, the whole premise collapses regardless
of the mechanism's elegance.  We measure three public multi-turn
trajectory corpora: Toolathlon-Trajectories~\cite{toolathlon},
CC-Bench-trajectories~\cite{ccbench}, and
hermes-agent-traces-filtered~\cite{hermesagent}.
Full-population pool: $7{,}530$ trajectories, $3.4{\times}10^8$
tokens.  Detailed analysis on a stratified subset of $1{,}035$
trajectories ($5.1{\times}10^7$ tokens).

We decompose each turn's tokens into \emph{prefix} (exact match,
served by SGLang today), \emph{PIC-cacheable} (same bytes at a shifted
position, our target), and \emph{novel} (new content) using 64-token
sliding-window xxHash~\cite{xxhash} fingerprints.

Across the three corpora, $12{-}24\%$ of unique content is
cross-session cacheable and within-session Toolathlon reaches p95
\textbf{48\%}; CC-Bench's lower rate ($4.6\%$ cross-session / $18.8\%$ within-session) correctly predicts
minimal PIC ROI for code-editing deployments, a workload-dependent
signal we preserve rather than average away.  These statistics measure
raw 64-token window repetition across heterogeneous trajectories --- a
lower bound on any single deployment.

\section{What Fixed-Block Hashing Misses}
\label{sec:algorithm}

\emph{Given that position-shifted reuse exists, can SGLang's current
fixed-block hash find it, and if not, how much is in principle
recoverable?}  This section answers the question \emph{offline} as a
structural probe; the runtime retrieval choice is made independently
in \S\ref{sec:algo}, and is \emph{not} the winning probe.

We benchmark three offline strategies.  \emph{Fixed-block} (SGLang
today) hashes every $B$-token aligned window --- a hit requires same
content at same offset.  \emph{CDC} (gear-hash) maintains an $O(1)$ Gear-hash rolling state
over a 64-token window and emits a chunk boundary when the rolling hash
falls below a threshold; each emitted chunk is then fingerprinted by a
single xxHash64 evaluation.  This produces variable-length chunks that
align to content regardless of byte offset.  \emph{CDC$+$fallback} adds a
second pass: on a CDC-chunk miss, sub-divide the chunk into fixed
$128$-token windows and check each sub-window hash.  All three are
$O(N)$ at single-digit ns/token.

Across Toolathlon, CC-Bench, and Hermes, fixed-block fails (agentic
messages start at heterogeneous byte offsets); CDC$+$fallback recovers
$3.8$--$4.8\times$ more across the two non-trivial corpora (Toolathlon
and Hermes), an \emph{upper bound} on what is structurally recoverable
from unannotated traffic.  The gap reflects a single underlying cost:
chunk boundaries that drift across requests when the rolling-hash state
sees different prefixes.  The runtime addresses this by
\emph{preventing} the drift at prompt-assembly time rather than
recovering at retrieval time; see \S\ref{sec:algo}.
\textbf{The table-winning offline strategy is a measurement probe,
not the deployed system.}

\section{Position-Invariance Across KV Components}
\label{sec:roc}

\emph{Even if we can find a reused chunk in the registry, we need to
ask: is the stored KV tensor actually safe to use at a new position?}
A content-invariant tensor is safe; a position-entangled tensor is
worse than useless --- it injects correlated errors.  We design a
threshold-free ROC test to answer this per-component.

\begin{figure}[t]\centering
  \includegraphics[width=\linewidth]{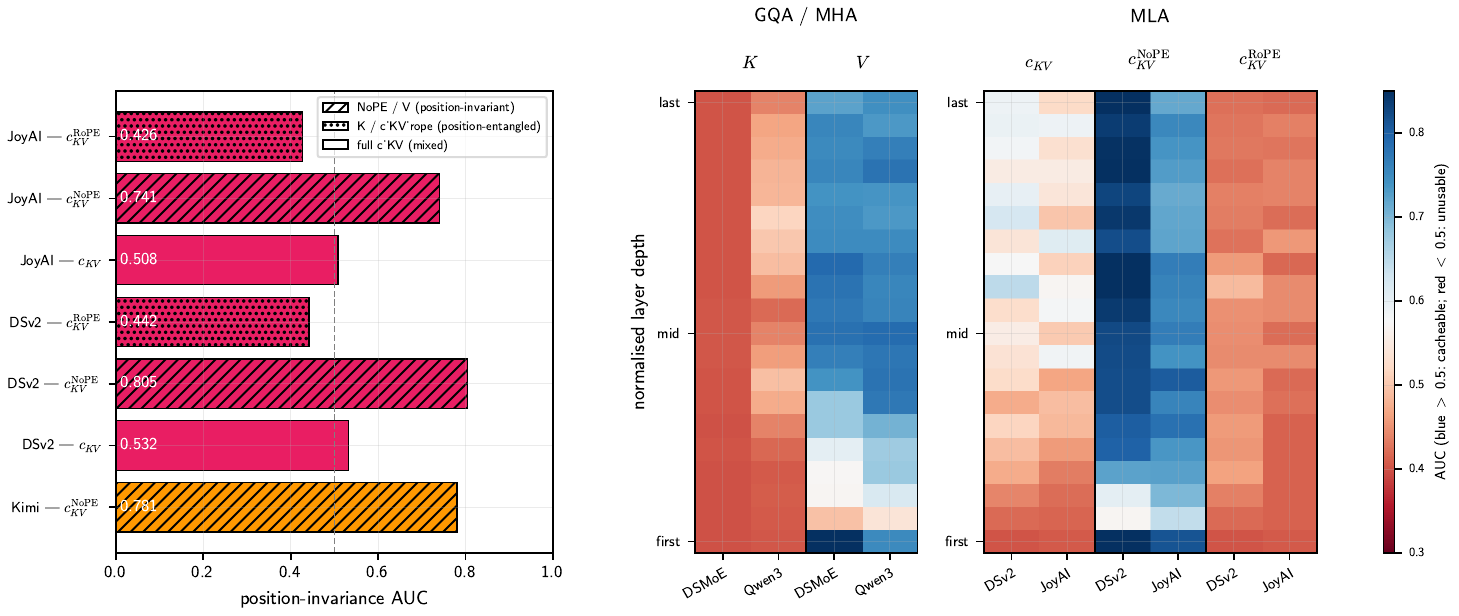}
  \caption{Position-invariance.
    \textbf{Notation:} the figure's
    $c_{KV}^{\mathrm{NoPE}}$ corresponds to the paper-text's $c_{KV}$
    (the $512$-dim position-free MLA latent, AUC$\approx\!0.77$);
    the figure's $c_{KV}^{\mathrm{RoPE}}$ corresponds to $k_r$
    ($64$-dim RoPE-rotated key, AUC$\approx\!0.43$); the figure's
    plain $c_{KV}$ label is a control showing the mixed / un-decoupled
    tensor (NoPE+RoPE concatenated as a single 576-dim row), which
    falls near AUC$\approx\!0.5$ as expected and is included to demonstrate
    that the structural split, not the latent itself, is what makes
    PIC viable.
    \textbf{Left:} AUC ranking for the MLA family.  Random line at 0.5.
    Hatching: diag=NoPE, dotted=RoPE.  Pure-softmax GQA/MHA omitted from
    the left panel: their $K$ falls below random and $V$-only reuse is
    insufficient without $K$ (see prose); both are shown on the right.
    \textbf{Right:} Per-layer AUC heatmap across all architectures.
    Inner-left: pure-softmax GQA/MHA --- $K$ below random (red), $V$
    above (blue).  Inner-right: MLA --- $c_{KV}^{\mathrm{NoPE}}$
    uniformly blue across every layer, $c_{KV}^{\mathrm{RoPE}}$
    uniformly red.}
\label{fig:roc-auc}
\end{figure}

We capture KV vectors from 500 content blocks placed at $5$ absolute
positions $\{0, 512, 1024, 2048, 3584\}$ in a 4\,096-token window;
within-block pairs (same content, different positions) are positives,
cross-block pairs (different content, randomised positions) are
negatives, and we sweep cosine-similarity threshold $\tau$ to compute
threshold-free AUC.  AUC$>0.5$ means content signal exceeds position
noise; AUC$<0.5$ means \emph{naive} reuse (inserting cached KV at a
new position without correction) is actively harmful.  The randomised
negative-position choice is the conservative one for a viability test
(Appendix~\ref{app:roc-methodology}).
Figure~\ref{fig:roc-auc} shows the AUC ranking for MLA's $c_{KV}$ and
$k_r$ (left) and the per-layer breakdown across all architectures (right).

The structurally-NoPE components win: MLA's $c_{KV}$ and KDA's
$c_{KV,\text{nope}}$ score AUC~$0.77$--$0.78$, cleanly separating
same-content-different-position from random cross-content.  V vectors
in every softmax architecture follow at AUC~$0.71$--$0.76$ (V is never
RoPE-rotated).  In our pure-softmax models, raw $K$ and MLA's $k_r$
score \emph{below random}: GQA~$K{=}0.453$ (Qwen3-32B),
MHA~$K{=}0.402$ (DS-MoE-16B), MLA~$k_r{=}0.432$.

The asymmetry is architectural.  GQA's \emph{entire} $K$ is below
random with no algebraic escape: any cache hit must pay correction
cost proportional to the full $d_K$.  MLA's $k_r$ is also below random
(AUC~$0.432$), but it is only $11\%$ of the KV row and is \emph{exactly}
correctable by $\delta$-rotation; the remaining $89\%$ ($c_{KV}$,
AUC~$0.77$) is directly reusable at zero per-hit cost.

Quantitatively, GQA's $K$ for Qwen3-32B
($8\text{ KV heads}{\times}128$ dims) is fully RoPE-entangled ---
$1{,}024$ dims of per-hit correction, $16{\times}$ MLA's $O(64N)$
$k_r$ rotation. This is the gap prior PIC systems mitigate but cannot close.

\section{Where Cache Hits Save Energy}
\label{sec:energy}

\emph{The ROC result tells us where PIC is mechanistically sound; the
energy result tells us what is at stake financially.}  We measure
\emph{existing} SGLang serving behaviour across three cache-event
classes (no Irminsul code in the loop): \emph{miss}, \emph{exact\_hit},
and \emph{partial\_recompute}$_k$.  NVML hardware energy counters
are sampled at event boundaries; 3 warm-up $+$ 10 timing repeats per
cell.

\begin{table}[t]
\centering\small
\caption{Prefill energy per (architecture, cache event) at
  seq\_len $4{,}096$.  ``Saving'' is $1{-}E_{\text{hit}}/E_{\text{miss}}$.
  $E_\text{partial}$ is the partial-recompute event class at $k{=}50\%$
  ($25\%$ and $75\%$ interpolate monotonically; supplemental).
  Cells aggregate non-zero NVML samples (the $\sim\!10$\,ms integration
  window can clip short small-model prefills); aggregation methodology
  and per-cell $n_{\text{non-zero}}{\geq}5$ in
  Appendix~\ref{app:scaffold}.}
\label{tab:energy}
\begin{tabular}{llrrrr}
\toprule
Architecture & Model & miss J & $E_\text{partial}$ J & hit J & Saving\\
\midrule
\textbf{GQA}    & Qwen3-32B~\cite{qwen3}                     & $262.3$ & $148.7$ & $37.5$ & $\mathbf{86\%}$\\
\textbf{MLA}    & DSv2-Lite~\cite{deepseekv2}\,+\,JoyAI-Flash$^{\ddagger}$~\cite{joyaillm} & $47.1$ & $43.9$ & $17.2$ & $\mathbf{63\%}$\\
\textbf{MHA}    & DS-MoE-16B$^*$~\cite{deepseekmoe}        & $45.2$  & $42.9$ & $15.3$ & $\mathbf{66\%}$\\
\midrule
Mamba2$^\dagger$ & Nemotron-3-Nano-30B-A3B~\cite{nemotron3} & $47.6$  & $47.7$ & $47.7$ & $\phantom{-}0\%$\\
GDN$^\dagger$    & Qwen3.6-35B-A3B~\cite{qwen3}              & $44.1$  & $43.9$ & $43.9$ & $\phantom{-}0\%$\\
KDA$^\dagger$    & Kimi-Linear-48B-A3B$^{\ddagger\ddagger}$~\cite{kdalinear}       & $37.5$  & $35.7$ & $37.4$ & $\phantom{-}0\%$\\
\bottomrule
\end{tabular}\\[2pt]
{\footnotesize
$^{*}$~Measured at seq\_len $2{,}048$, the model's context limit.
$^{\dagger}$~Hybrid: SSM/linear layers interleaved with a minority of
softmax-attention layers.  The linear/SSM state neither appends nor
forks, so every request re-runs the full recurrent pass regardless of
cache state~\cite{marconi}, so $E_\text{partial}{\approx}E_\text{miss}$
by construction.
$^{\ddagger}$~JoyAI-LLM-Flash is a $48$B-total/$3$B-active MLA-MoE
(\texttt{jdopensource/JoyAI-LLM-Flash}).  The MLA row reports the per-cell
mean over DSv2-Lite and JoyAI-Flash.
$^{\ddagger\ddagger}$~Kimi-Linear-48B-A3B
(\texttt{moonshotai/Kimi-Linear-48B-A3B-Instruct}) is Moonshot AI's
KDA-hybrid (linear-attention + MLA-NoPE, $3{:}1$ ratio); architecturally
distinct from \emph{Moonlight-16B-A3B}~\cite{moonlight}
(\texttt{moonshotai/Moonlight-16B-A3B-Instruct}, native MLA-MoE in DSv3
form), which appears in the output-consistency study
(Table~\ref{tab:consistency}).  The two are siblings only by org.}
\end{table}

A prefix hit costs $14$/$37$/$34\%$ of miss energy on GQA/MLA/MHA,
a $2.6$--$7\times$ reduction.  The $50\%$ partial-recompute class lands
cleanly between miss and hit on Qwen3-32B (the cell with the largest
absolute energy budget), but sits close to miss on the smaller
MLA/MHA cells: at the model scales of DSv2-Lite, JoyAI, and DS-MoE-16B,
fixed per-prefill kernel-launch and routing costs dominate the savings
that 50\%-KV reuse can elide, leaving the headline benefit
concentrated in the full-hit case.
Hybrid Mamba2, GDN, and KDA show
\emph{zero} savings across three independent families and an
$E_\text{partial}{\approx}E_\text{miss}$ relationship: their recurrent
state is not token-indexed, so ``cache hit'' is not a defined operation
at the SSM layers (\S\ref{sec:roc}).  This bounds where any PIC system,
ours included, can deliver value.

\paragraph{Scope: absorbed-matmul MLA.}
Irminsul's energy and recovery claims target the absorbed-matmul form
of MLA --- the form used in every production MLA deployment we are
aware of (DSv2/V3/R1, Kimi-K2/Moonlight, JoyAI-Flash etc.) --- in which $W_{uk}$ is folded into the query projection so
attention executes directly on the $512$-dim $c_{KV}$.  The only
public dense MLA model we are aware of, MiniCPM3-4B, uses a
non-absorbed runtime that up-projects $c_{KV}$ to a full-$d_K$ key per
token before attention; the per-hit compute then tracks GQA, and the
energy curves of Table~\ref{tab:energy} would not transfer.  The two
production MLA deployments studied here (DSv2-Lite, JoyAI-Flash) are
both MLA-MoE in absorbed form, which we believe is the only
configuration in current production use; the convergence is not
incidental --- absorbed-matmul is the property that makes MLA's
storage advantage translate into a serving advantage, and is therefore
what content-addressed caching exploits.

\section{Irminsul}
\label{sec:algo}

\emph{We now have all the pieces: reuse exists (\S\ref{sec:workload}),
CDC finds it (\S\ref{sec:algorithm}), MLA's $c_{KV}$ is safe to reuse
(\S\ref{sec:roc}), and a hit saves substantial prefill energy
(\S\ref{sec:energy}).}  One remaining design question is what to do
with the first chunk, where the attention-sink
phenomenon~\cite{streamingllm} concentrates: EPIC recomputes its
boundary tokens for this reason.

\subsection{First-Chunk Carve-Out}
\label{sec:sink}

\textbf{Attention sink is sequence-start-local, not chunk-start-local.}
We measure the fraction of intra-chunk attention absorbed by the first
$k{=}32$ tokens of a chunk placed at absolute position $p$.
If sinks are \emph{chunk-start-local} (EPIC's implicit assumption),
every chunk placed at any position would need boundary recompute.
If sinks are \emph{sequence-start-local}, only position-0 chunks matter.

\begin{figure}[h]\centering
  \includegraphics[width=0.99\linewidth]{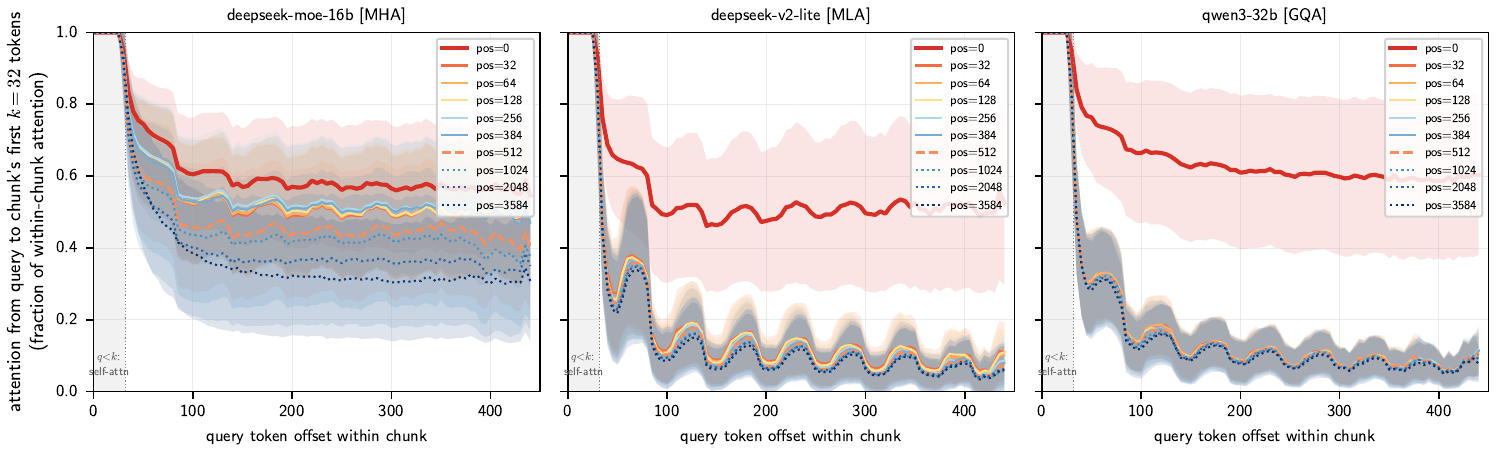}
  \caption{Sink ratio at $k{=}32$ vs.\ intra-chunk query offset, one
    panel per architecture.  Colour encodes chunk absolute position
    (red~$p{=}0$; warm gradient $p{\in}\{32,\ldots,384\}$; cool blue
    $p{\geq}512$).  Grey band $q{<}32$: trivial regime (ratio${=}1$ by
    causal masking; read only $q{\geq}32$).  MLA and GQA: every
    $p{\geq}32$ curve collapses to the $p{\geq}1024$ baseline in a
    one-shot step, justifying the $p{=}0$ carve-out.  MHA: elevated
    through $p{\sim}384$.  Bands ${\pm}1$\,s.d.\ (20 samples).}
  \label{fig:sink}
\end{figure}

Figure~\ref{fig:sink} confirms the result.  At $p{=}32$ the sink ratio
has already collapsed to $0.258$ (MLA, $p{\geq}1024$ baseline $0.230$)
and $0.251$ (GQA, baseline $0.240$), within noise of the floor,
while MHA stays at $0.589$, a qualitatively different shape.  A
carve-out at $p{=}0$ is therefore sufficient for MLA.

A \textbf{first-chunk carve-out} --- prefilling all tokens at $p{<}32$
via ordinary prefill (Algorithm~\ref{alg:irminsul}, line~5) --- therefore
sidesteps the entire sink regime without any chunk-boundary recompute on
later chunks. CDC's $\geq\!32$-token chunk-size clamp guarantees this is
the worst case for any chunk Irminsul serves.
A direct measurement of the carve-out under PIC --- tail-attention mass
to absolute $[0, 32)$ on PIC-displaced bodies, three models --- shows
PIC preserves the sink-anchor mass within mean absolute deviation
$\leq\!0.003$ of fresh prefill, while naive reuse drifts $2.4{-}9.0{\times}$
farther.

A complementary metric --- the absolute $[0,32)$ probability mass S
(how much of a chunk-token's total softmax attention reaches back to
the true sequence start, regardless of chunk position) --- shows that
for GQA and MLA the sink remains anchored at the sequence start out to
$p{=}3584$ (S$\approx 0.4{-}0.6$) while the within-chunk fraction W
stays flat near $0.25$; for MHA the pattern reverses, suggesting a
locally-concentrated sink (Appendix~\ref{app:sink-anchor}).
For GQA and our deployment target MLA, chunk~$0$ covers
absolute $[0,32)$ by construction, anchoring later chunks'
globally-pulled sink attention.  MHA's locally-concentrated sink lies
outside this mechanism, so we do not claim the carve-out alone handles
MHA.  Our consistency harness in \S\ref{sec:precision} is MLA-specific
(it depends on the $64$-dim $k_r$ slice for $\delta$-rotation), so we do
not have an MHA cell in Table~\ref{tab:consistency} either; an MHA-PIC
study would require either an EPIC-style chunk-boundary recompute path
or a different KV factoring, and we leave that to follow-up.  Within
this paper, the load-bearing safety claim covers MLA only; MHA appears
solely in the energy and ROC studies for cross-architecture context.

\subsection{Algorithm}
Algorithm~\ref{alg:irminsul} sketches the hot path.  Chunk-boundary
drift is prevented at prompt-assembly time by a $64$-token CDC
boundary marker around each shared region (S2,
\emph{prevent-via-cooperation}); when the assembler is not under operator
control, the offline CDC$+$fallback of \S\ref{sec:algorithm}
(S1, \emph{detect-and-recover}) is the deployment-time fallback. S1 and
S2 compose; details in Appendix~\ref{app:strategies}.

\begin{algorithm}[h]
\small
\caption{Irminsul serve path (per incoming prompt).}
\label{alg:irminsul}
\begin{algorithmic}[1]
\Require prompt tokens $T$;\ registry $\mathcal{R}:\ \text{hash}\mapsto(c_{KV},\ k_{r,\text{base}},\ p_{\text{src}})$
\Statex \textsc{Notation:} $s$ = segment (chunk of tokens);\ $p$ = absolute position of $s$ in $T$;\ $\delta = p - p_{\text{src}}$.
\State $\textit{prefix},\ \textit{tail} \gets \textsc{PrefixMatch}(T)$ \Comment{standard RadixCache exact-prefix hit}
\State $\textit{segments} \gets \textsc{CDC}(\textit{tail})$ \Comment{Gear-hash boundaries; xxHash64 fingerprints each emitted chunk}
\State $\textit{kv\_out} \gets \textit{prefix}.\textit{kv}$
\For{each segment $s$ in $\textit{segments}$ at absolute position $p$}
  \If{$p < 32$} \Comment{first-chunk carve-out: sink regime (\S\ref{sec:algo})}
    \State $\textit{kv} \gets \textsc{Prefill}(s)$
  \ElsIf{$\mathcal{R}[\textsc{xxhash}(s)]$ \textbf{hits} as $(c_{KV},\ k_{r,\text{base}},\ p_{\text{src}})$}
    \State $\delta \gets p - p_{\text{src}}$
    \State $k_r \gets R(\delta)\cdot k_{r,\text{base}}$ \Comment{$R(\delta)R(p_{\text{src}}){=}R(p)$, fused in FlashMLA}
    \State $\textit{kv} \gets (c_{KV},\ k_r)$
  \Else
    \State $\textit{kv} \gets (c_{KV},\ k_{r,\text{base}}) \gets \textsc{Prefill}(s)$ \Comment{novel content under S2; S1 is an extension (\S\ref{sec:algorithm}, \S\ref{sec:algo})}
    \State $\mathcal{R}[\textsc{xxhash}(s)] \gets (c_{KV},\ k_{r,\text{base}},\ p)$ \Comment{insert into registry for future hits}
  \EndIf
  \State $\textit{kv\_out}.\textsc{append}(\textit{kv})$
\EndFor
\State \Return $\textit{kv\_out}$
\end{algorithmic}
\end{algorithm}

The \textbf{CDC boundary rule} (Gear-hash rolling state, low-$7$-bit
mask, expected chunk $\sim\!128$ tokens clamped to $[32,512]$,
xxHash64 fingerprint per chunk) and the mask-exponent ablation
selecting $k{=}7$ are in Appendix~\ref{app:cdc-mask}.

\textbf{$k_r$ rotation} applies $R(\delta)$ uniformly across every
token (Algorithm~\ref{alg:irminsul}, line~9) --- $O(N{\cdot}64)$
multiplies, fuseable into FlashMLA's HBM$\to$SRAM load path with no
extra HBM bandwidth.  The runtime layout (split TokenToKVPool with a
content-hashed $c_{KV}$ pool shared across sessions and a per-request
$k_r$ pool, plus the gather--rotate--scatter materialisation in our
scaffold vs.\ a fused-load production variant) is described in
Appendix~\ref{app:runtime-layout}.

\subsection{Quality and the RoPE Pitfall}
\label{sec:precision}

The algebraic $\delta$-rotation error is $4.7{\times}10^{-3}$ rel-L2 in
bf16, non-accumulating, an order of magnitude tighter than production
FP8 KV quantisation ($10^{-2}$--$3{\times}10^{-2}$), which is accepted
practice.  On \textsc{hotpotqa}$\cup$\textsc{musique} across DSv2-Lite,
TransMLA-4B, and JoyAI-48B, Irminsul's task F1 stays within Wilson SEM
of full recompute, but absolute F1 on these sparse extraction tasks is
too small to discriminate cleanly between PIC and naive reuse; F1 is
therefore a coarse safety check, and the load-bearing evidence is the
output-consistency study below, including Moonlight-16B-A3B as an additional configuration, in Table~\ref{tab:consistency}.

\textbf{Output-consistency across models and datasets.}
Beyond F1, we teacher-force full prefill's greedy trajectory through
each cache path and measure per-token KL$(p_{\text{full}}\|p_x)$,
$\mathrm{argmax}$-match, and free-running first-divergence position
on three native MLA-MoE deployments (DSv2-Lite $16$B/$2.4$B;
Moonlight-16B-A3B~\cite{moonlight}; JoyAI-Flash
$48$B/$3$B; load-bearing) plus a retrofit stress-test
(TransMLA-4B$^{\#}$).
Across all four datasets (QA, govreport, NIAH; $n{=}30{-}100$ per
cell), PIC matches or beats naive reuse on every populated cell except
DSv2-Lite \textsc{govreport} ($\pm0.02$ KL); PIC's greedy trajectory
stays identical to full prefill for up to $2.4{\times}$ more tokens
before divergence; on JoyAI-48B NIAH, needle recall matches full
prefill exactly ($0.820$; naive $0.840$ within noise) while KL drops
$22\%$ vs.\ naive.  Methodology, per-cell discussion, and
TransMLA-4B's NIAH omission rationale (full-prefill recall is $0\%$,
so the cell would not discriminate) are in
Appendix~\ref{app:consistency}.

\begin{table}[h]\centering\small
\caption{Output-consistency. QA = pooled means over
  \textsc{hotpotqa}$\cup$\textsc{musique}; \textsc{govrep} =
  free-form summarisation (GovReport); \textsc{niah} = synthetic
  needle-in-a-haystack recall.  KL$_x$: per-step
  KL$(p_\text{full}\|p_x)$.  AM$_x$: $\mathrm{argmax}$-match rate
  vs.\ full prefill (1 = identical token at every step).
  $\Delta_x$: mean first-divergence position in greedy decoding.
  $^{\S}$~TransMLA-4B's base needle recall on this NIAH configuration
  is $0\%$ (full-prefill itself fails to retrieve the needle), so the
  cell would not discriminate PIC from naive reuse and is omitted.
  $^{\#}$~TransMLA-4B is a Minitron-4B checkpoint retrofitted to MLA
  via post-hoc KV factoring; long-context recall is degraded, and
  QA / govrep cells are reported within the model's effective context.
  Its primary role is the rotary-pitfall study
  (Appendix~\ref{app:rope-pitfall}).}
\label{tab:consistency}
\begin{tabular}{llrrrrrr}
\toprule
Model & Dataset & KL$_\text{pic}$ & KL$_\text{naive}$ & AM$_\text{pic}$ & AM$_\text{naive}$ & $\Delta_\text{pic}$ & $\Delta_\text{naive}$ \\
\midrule
\multirow{4}{*}{TransMLA-4B$^{\#}$ (retrofit)}
       & QA              & $\mathbf{0.054}$ & $0.112$ & $\mathbf{0.885}$ & $0.852$ & $\mathbf{6.5}$ & $4.5$ \\
       & \textsc{govrep} & $\mathbf{0.173}$ & $0.175$ & $\mathbf{0.841}$ & $0.829$ & $\mathbf{3.9}$ & $2.8$ \\
       & \textsc{niah}$^{\S}$   & --- & --- & --- & --- & --- & --- \\
\midrule
\multirow{4}{*}{DSv2-Lite ($16$B/$2.4$B native MoE)}
       & QA              & $\mathbf{0.146}$ & $0.210$ & $\mathbf{0.904}$ & $0.876$ & $\mathbf{5.8}$ & $3.2$ \\
       & \textsc{govrep} & $0.297$ & $\mathbf{0.276}$ & $0.922$ & $\mathbf{0.927}$ & $\mathbf{9.5}$ & $11.0$ \\
       & \textsc{niah}   & $\mathbf{0.053}$ & $0.063$ & $\mathbf{0.936}$ & $0.933$ & $\mathbf{9.7}$ & $9.0$ \\
\midrule
\multirow{4}{*}{Moonlight-16B ($16$B/$3$B native MoE)}
       & QA              & $\mathbf{0.051}$ & $0.070$ & $\mathbf{0.937}$ & $0.930$ & $\mathbf{9.9}$ & $9.0$ \\
       & \textsc{govrep} & $\mathbf{0.116}$ & $0.137$ & $\mathbf{0.903}$ & $0.897$ & $\mathbf{12.0}$ & $11.0$ \\
       & \textsc{niah}   & $\mathbf{0.021}$ & $0.031$ & $0.977$ & $0.977$ & $\mathbf{25.2}$ & $24.8$ \\
\midrule
\multirow{4}{*}{JoyAI-48B ($48$B/$3$B native MoE)}
       & QA              & $\mathbf{0.539}$ & $0.743$ & $\mathbf{0.782}$ & $0.754$ & $\mathbf{1.7}$ & $0.7$ \\
       & \textsc{govrep} & $\mathbf{0.369}$ & $0.419$ & $\mathbf{0.793}$ & $0.768$ & $\mathbf{2.9}$ & $2.2$ \\
       & \textsc{niah}   & $\mathbf{0.484}$ & $0.624$ & $\mathbf{0.823}$ & $0.744$ & $\mathbf{5.2}$ & $2.7$ \\
\bottomrule
\end{tabular}
\end{table}

A mismatched positional frequency $\theta$ produces a silently-wrong
rotation that collapses PIC below naive reuse; the $\delta$-rotation
is therefore architecturally load-bearing, and the model's own rotary
class must be auto-detected.  The collapse-and-recovery case study,
with numbers, is in Appendix~\ref{app:rope-pitfall}.

\subsection{Recovery Measurement and the Partition-Shift Diagnostic}
\label{sec:ab}

We measure on a 950-LoC reference scaffold running in observer mode
(every would-be hit recorded without rerouting KV), so Table~\ref{tab:ab}
reports recovery rates under treatment and baseline TTFT only;
end-to-end treatment TTFT requires the FlashMLA-fused $\delta$-rotation
kernel and is a deliberate follow-up.  The marker-pinned S2 path is
load-bearing: removing the 64-token boundary marker collapses
content-hash matching from the table's ${\sim}77\%$ to ${\sim}1\%$.
Scaffold mechanics, observer-mode invariants, and the marker control
are detailed in Appendix~\ref{app:scaffold}.

\begin{table}[h]\centering\small
\caption{Irminsul-unique token recovery \emph{above and beyond}
  SGLang exact-prefix.  Default: DeepSeek-V2-Lite; agent\_meta also
  measured on JoyAI-48B for cross-model confirmation.
  Flagged runs: $n_{\text{req}}{=}160$; others $n{=}80$.}
\label{tab:ab}
\begin{tabular}{llrrrr}
\toprule
Pattern & Model & tprefix$^a$ & \textbf{pic-unique}$^b$ & total$^c$ & TTFT (ms)$^d$ \\
\midrule
agent\_meta$^\dagger$ & DSv2-Lite & $1.9\%$  & $\mathbf{77.2\%}$ & $79.1\%$ & $325$ \\
agent\_meta           & JoyAI-48B & $1.5\%$  & $\mathbf{82.7\%}$ & $84.3\%$ & $582$ \\
sysvar$^\dagger$      & DSv2-Lite & $73.7\%$ & $26.3\%$          & $100\%$  & $302$ \\
compact               & DSv2-Lite & $99.5\%$ & $0.4\%$           & $99.9\%$ & $268$ \\
rerank                & DSv2-Lite & $99.9\%$ & $0.04\%$          & $99.9\%$ & $274$ \\
tool\_variants        & DSv2-Lite & $99.96\%$& $0.04\%$          & $100\%$  & $283$ \\
\bottomrule
\end{tabular}\\[2pt]
{\footnotesize
$^{\dagger}$~$n_{\text{req}}{=}160$.
$^a$~Fraction of prompt tokens served by SGLang's exact-prefix cache.
$^b$~Additional fraction Irminsul recovers above exact-prefix (the headline metric).
$^c$~$a{+}b$: total tokens served from any cache, not re-prefilled.
$^d$~Median time-to-first-token for the \emph{baseline} (observer mode).  Treatment TTFT is intentionally not reported here (see prose: requires the fused $\delta$-rotation kernel that is out of scope for the scaffold).
}
\end{table}

The results partition cleanly by where variation enters the prompt. On \texttt{agent\_meta}, where a small per-agent header shifts the
bulk of the prompt, Irminsul recovers $77.2\%$ and exact-prefix gets
$1.9\%$; together they serve $79.1\%$ from cache.  On \texttt{sysvar},
\texttt{compact}, \texttt{rerank}, and \texttt{tool\_variants}, the
variation occurs deeper in the prompt, so exact-prefix already captures
$74{-}100\%$, leaving little for Irminsul.
This is the partition-shift diagnostic: \textbf{Irminsul's marginal
value equals the fraction of tokens shifted beyond the first variation
point.}  Operators can predict Irminsul ROI from
two measurable statistics: first-variation position and
variation-set cardinality.

\begin{figure*}[t]\centering
\includegraphics[width=0.92\textwidth]{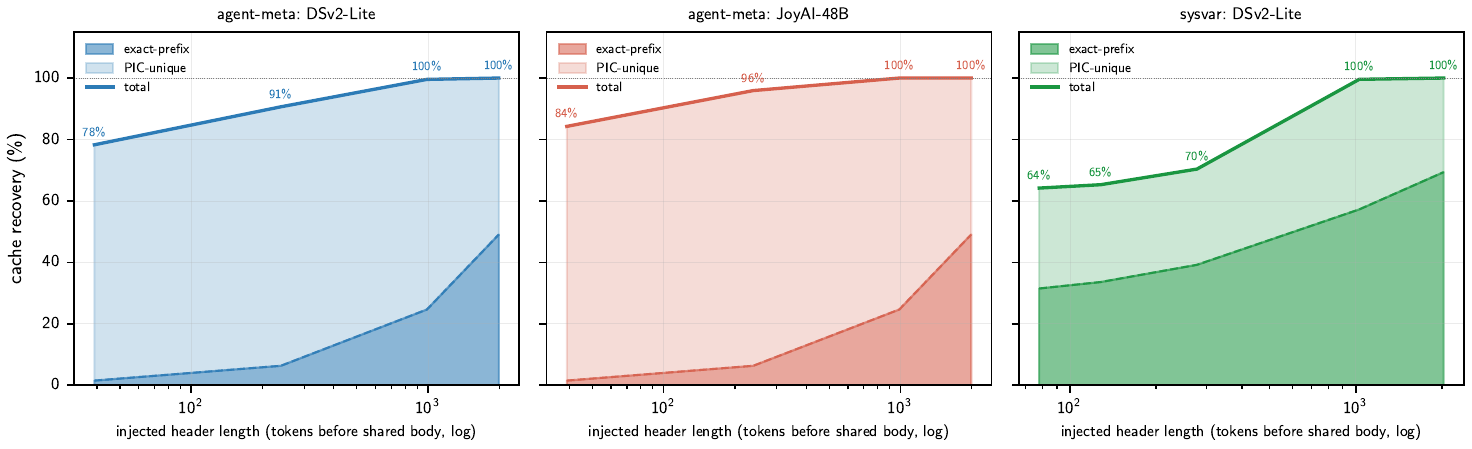}
\caption{\textbf{Partition-shift: PIC and exact-prefix occupy complementary
  regimes.}
  Light fill = exact-prefix; hatched = Irminsul-unique; totals annotated;
  dashed = 100\% ceiling.
  \textbf{Left:} \texttt{agent\_meta} (early-variation): variation near
  prompt start shifts most tokens, so PIC dominates.  Two curves:
  DSv2-Lite (blue, stacked area) and JoyAI-48B (orange, line+marker
  total) sweep the same four header lengths $\{50, 250, 1000, 2000\}$;
  JoyAI tracks DSv2 within $0$--$10$ pp at every point despite
  $3{\times}$ the parameter scale, confirming the cross-model claim.
  \textbf{Right:} \texttt{sysvar} (deep-variation): shared prefix deepens
  with header size, so exact-prefix grows and PIC fills the residual gap
  (DSv2-Lite only).}
\label{fig:partition-shift}
\end{figure*}

On agent\_meta, DSv2-Lite and JoyAI-48B recover $77.2\%$ vs $82.7\%$
above exact-prefix --- both in the high-$70\%$ to low-$80\%$ band
despite $3\times$ parameter scale --- Irminsul's value is fixed by
the workload's shift structure, not the model.  Figure~\ref{fig:partition-shift} sweeps the injected header
length (tokens inserted \emph{before} the shared body) and confirms
total recovery climbs monotonically with header size: on
\texttt{agent\_meta} (early-variation), exact-prefix grows from
$1.5\%$ to $49.1\%$ as a longer shared header lengthens the prefix,
Irminsul-unique fills the residual shifted body, and the
sum saturates at $100\%$; on \texttt{sysvar} (deep-variation), the
shared prefix lengthens with the header so exact-prefix grows and
Irminsul keeps the residual gap closed, also saturating at $100\%$.

\section{Conclusion}

Content-addressed caching, on the analysis here, is less a caching
trick than a recognition that agentic prompts compose at runtime and
the serving stack should index KV by what the tokens \emph{are}, not
where they sit.  Two structural axes, not one, govern where PIC pays off.  The
position-invariance ROC (\S\ref{sec:roc}) shows that a softmax
architecture admits content-addressable reuse \emph{at zero per-hit
correction cost} only to the extent its KV factors out position: MLA
does, GQA does not, and the gap is structural rather than algorithmic, though GQA still admits PIC at full-$d_K$ correction cost as prior
work demonstrates.  Orthogonally, the cross-architecture energy study
(\S\ref{sec:energy}) shows that whether a cache hit saves $63{-}86\%$
of prefill energy or essentially zero is decided by softmax-vs-recurrent
attention, not position factoring: pure-softmax GQA, MLA, and MHA all
realise large savings on a hit, while hybrid SSM/linear architectures
re-run the recurrent pass regardless and save ${\approx}0\%$.  (The
safety claim covers MLA only; MHA's locally-concentrated sink would
require an EPIC-style boundary recompute path we do not study --- see
\S\ref{sec:sink}.)  MLA's
distinct contribution is reducing the per-hit correction cost to
$O(64N)$, making PIC deployable inside production latency budgets, not
enabling cache-hit energy savings that GQA already obtains.

The agentic era is intensifying that pressure in a specific way. Each turn of a multi-agent pipeline re-encodes shared context that was already paid for: tool schemas re-read, retrieved documents re-prefilled, system prompts re-attended. The serving stack is currently billing that redundancy as \emph{compute}; Irminsul's central claim is that it is \emph{memory}. At the scale of a production agentic deployment --- thousands of concurrent sessions, each replaying kilobytes of shared context every few seconds --- the difference is not academic. Token re-encoding is the new idle spin.

The architecture community has arrived at the same diagnosis from the model side. DeepSeek-V4's Heavily Compressed Attention~\cite{deepseekv4} reflects a recognition that per-token KV footprints, when multiplied across the long contexts and large batch sizes that agentic workloads demand, become the binding constraint, and responds by compressing that footprint at the representational level. The motivation converges with Irminsul's: if KV is the bottleneck, reduce it. The intervention point differs: DSv4 reshapes the representation before it is stored; Irminsul avoids storing it redundantly in the first place. The two approaches are orthogonal, but HCA's aggregated token identity dissolves the per-token content-hash that Irminsul's chunking primitive relies on. Irminsul does not run on DSv4 HCA layers as-is, and adapting the chunking primitive to HCA's aggregation schedule is the most immediate follow-up (Appendix~\ref{app:hca}). 

The question facing the next generation of agentic-serving systems is not whether to compress KV, nor whether to cache it --- both are settled --- but whether the representation is factored cleanly enough that tokens, once paid for, need never be paid for again.

\newpage

{\small
\bibliographystyle{plainnat}
\bibliography{references}
}

\newpage

\appendix

\section{Sink Anchoring: W vs.\ S}
\label{app:sink-anchor}

Table~\ref{tab:sink-anchor} pairs two metrics at $k{=}32$.  The
within-chunk fraction W is what the carve-out decision in
\S\ref{sec:sink} rested on.  A complementary question is the absolute
$[0,32)$ probability mass S: how much of a chunk-token's total softmax
attention lands on the true sequence start, regardless of where the
chunk is placed.

\begin{table}[h]\centering\small
\caption{Two anchor metrics for the attention sink at $k{=}32$, by
  chunk position.  W: within-chunk first-$32$ attention fraction.
  S: absolute $[0,32)$ probability mass (sequence-start fraction).
  At $p{=}0$ both metrics coincide by construction.}
\label{tab:sink-anchor}
\begin{tabular}{llcccccc}
\toprule
& & \multicolumn{2}{c}{$p{=}32$} & \multicolumn{2}{c}{$p{=}512$} &
  \multicolumn{2}{c}{$p{=}3584$} \\
\cmidrule(lr){3-4}\cmidrule(lr){5-6}\cmidrule(lr){7-8}
Arch & Model & W & S & W & S & W & S \\
\midrule
GQA & Qwen3-32B  & $0.25$ & $\mathbf{0.61}$ & $0.25$ & $\mathbf{0.55}$ & $0.24$ & $\mathbf{0.50}$ \\
MLA & DSv2-Lite  & $0.26$ & $\mathbf{0.48}$ & $0.24$ & $\mathbf{0.43}$ & $0.23$ & $\mathbf{0.42}$ \\
MHA & DS-MoE-16B  & $\mathbf{0.59}$ & $0.23$          & $\mathbf{0.54}$ & $0.06$ & $\mathbf{0.44}$ & $0.008$ \\
\bottomrule
\end{tabular}
\end{table}

\begin{figure}[h]\centering
  \includegraphics[width=0.99\linewidth]{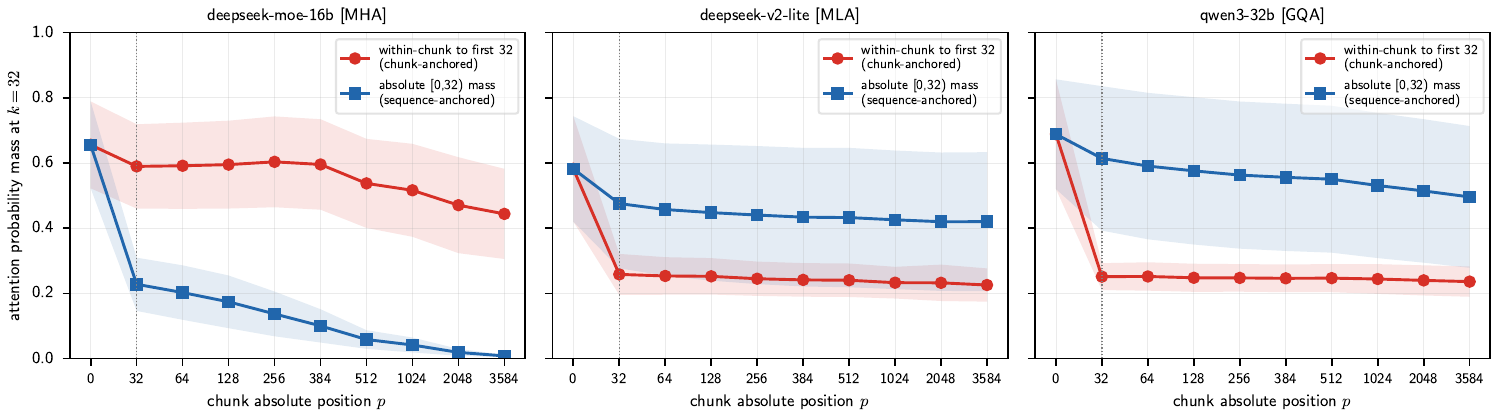}
  \caption{Sink anchoring across architectures (Figure~\ref{fig:sink}
    complement).  Red: within-chunk first-$32$ fraction~(W).
    Blue: absolute $[0,32)$ probability mass~(S).  Both coincide at
    $p{=}0$.  For MLA and GQA, S remains near $0.4{-}0.6$ while W
    stays near $0.25$; for MHA the pattern reverses.
    ${\pm}1$\,s.d.\ across layers.}
  \label{fig:sink-anchor}
\end{figure}

For GQA and MLA, W stays near $0.25$ at all positions while S holds
near $0.4{-}0.6$ out to $p{=}3584$: a substantial fraction of each
chunk's attention budget reaches back to the absolute sequence start
regardless of where the chunk sits.  MHA shows the opposite pattern: W
remains elevated while S decays to ${\approx}0$, suggesting the sink
concentrates locally within the chunk rather than at the global start.
We note this as an observation on three model instances, not a
conclusion about the architectures more broadly.

\paragraph{Why position 0 (and not some other position) is the sink.}
Recent theoretical work explains the architectural asymmetry empirically
visible above.  Ran-Milo et al.~\cite{ranmilo2026sink} prove that
attention sinks are a necessary feature of softmax-attention transformers
trained with autoregressive next-token-prediction loss: the softmax
denominator forbids true ``no-attention'', so escape mass must
concentrate somewhere, and three structural facts pin it to
position~0 in the GQA/MLA regime.
\emph{(i)~Universal visibility under causal masking.}  Position~0 is
the only token visible to every later query; this enables the
content-free identification circuit
(\emph{P0-Sink Circuit}~\cite{p0sink2026}) that routes escape mass to
position~0 without requiring any semantic feature.
\emph{(ii)~No prediction-loss pressure on position~0.}  Position~0 is
the only token never the target of an autoregressive prediction step,
so no gradient pressure penalises its use as an attention dump.
\emph{(iii)~Self-reinforcing gradient sinks.}  Once a position
absorbs escape mass, the same gradient mechanism that makes it a
forward-pass sink also makes it a \emph{gradient sink} that suppresses
updates pulling away from it~\cite{chen2026gradsink}, so the
configuration is locally stable under continued training.  A
graph-theoretic formalisation of the same position-0 privilege from
the perspective of position bias is given by~\cite{posbias2025}.
The condition for these arguments to predict a \emph{single} dominant
sink at position~0 is that escape mass concentrates at one position
($\arg\max_p \mathbb{E}[a_{q\to p}]$ unique across heads); when
multi-head dynamics distribute escape mass across several positions
(MHA with unconverged head specialisation in our DS-MoE-16B
measurement), the prediction reduces to the empirical pattern of
Table~\ref{tab:sink-anchor}'s third row, where W dominates S.
We treat this connection as an explanatory lens for the empirical
result rather than a derivation of it.

\section{CDC Mask-Exponent Ablation}
\label{app:cdc-mask}

The Gear-hash boundary mask exponent $k$ controls expected chunk size
($\sim\!2^k$ tokens, clamped to $[32,512]$).  A mask-exponent sweep
across the three corpora of \S\ref{sec:algorithm} shows $k{=}7$ gives
$1.3{-}2.3{\times}$ the offline dedup rate of $k{=}16$ at identical
max\_size.  The direction is intuitive: a wider mask produces
expected chunk lengths above max\_size, so chunks clamp at max\_size
and degenerate to fixed-block hashing --- the failure mode the
content-hash design is meant to escape.  $k{=}7$ is the production
default.

\section{Runtime KV Layout}
\label{app:runtime-layout}

\paragraph{$k_r$ rotation materialisation.}
In our reference scaffold we materialise the rotated $k_r$ tensor into
the per-request pool slot in HBM at retrieval time (an in-place
gather--rotate--scatter against the destination indices), so the
registry stores only the canonical $k_{r,\text{base}}$ once per content
hash and per-request HBM holds each request's already-rotated copy.
A production deployment can instead keep $k_{r,\text{base}}$ in HBM
and fuse the rotation into the FlashMLA load path so the rotated
tensor never materialises; both schemes share the registry and differ
only in where the rotation lands.

\paragraph{Split TokenToKVPool.}
The split TokenToKVPool partitions the $576$-dim KV row into a
content-hashed shared $c_{KV}$ pool ($512$ dim) and a per-request $k_r$
pool ($64$ dim), so position-shifted shared content (e.g.\ a
dynamically-placed tool schema, a retrieved document, or a system
prompt with prepended per-agent metadata) deduplicates its $89\%$
$c_{KV}$ payload in HBM across concurrent sessions; only the $11\%$
position-carrying $k_r$ slice is materialised per request.
Sequence-start system prompts that are byte-identical across sessions
are already covered by the standard exact-prefix path
(Algorithm~\ref{alg:irminsul}, line~1); the split pool's distinctive
utility is precisely the position-shifted case.

\section{RoPE-Frequency Pitfall: A Case Study}
\label{app:rope-pitfall}

A mismatched positional frequency $\theta$ produces a silently-wrong
rotation.  In our consistency harness, TransMLA-4B run with the
obvious plain-RoPE fallback (base${=}10000$) collapses PIC to
argmax-match $0.125$ and KL${=}3.51$, worse than naive reuse on the
same workload.  With the model's own rotary class auto-detected
(\texttt{DeepseekV3RotaryEmbedding} + \texttt{attention\_scaling}
de-duplication), PIC recovers to argmax-match $0.888$ and KL${=}0.054$
(matching Table~\ref{tab:consistency}'s TransMLA-QA cell).  The
$\delta$-rotation is architecturally load-bearing, not decorative; when
implemented against the model's actual rotary, the absence of collapse
on four models (DSv2-Lite, Moonlight-16B, JoyAI-48B, TransMLA-4B,
spanning DSv2-form and DSv3-form rotaries and $\theta$ values from
$10^4$ to $3.2{\times}10^7$) is positive evidence rather than a
default outcome.

\paragraph{Implementation note: deriving cos/sin from \texttt{inv\_freq}.}
We compute the $\delta$-rotation cos/sin directly from the rotary
module's \texttt{inv\_freq} buffer rather than calling
\texttt{rotary\_emb(\,$\cdot$\,, position\_ids)}.  The latter triggers a
\texttt{torch.arange(Tensor)} regression in transformers~4.57 on
\texttt{DeepseekV3RotaryEmbedding} that throws and silently falls
through to a plain-RoPE base${=}10000$ fallback --- the exact failure
mode this case study warns about.  Reading \texttt{inv\_freq} directly
recovers the model's true $\theta$ (e.g.\ JoyAI-48B's $\theta{=}3.2{\times}10^7$,
Moonlight's $\theta{=}5{\times}10^4$) without depending on the
upstream call path.  Auditing this on each new model is a one-line
check against a hand-computed reference; we recommend any
$\delta$-rotation implementation include it.

\section{Reference Scaffold and Recovery-Measurement Methodology}
\label{app:scaffold}

The 950-LoC reference scaffold patches \texttt{RadixCache} via a
subprocess-safe \texttt{sitecustomize} hook.  In observer mode
(\texttt{IRMINSUL\_PIC\_LIVE=1}) we record every would-be hit without
rerouting KV, so measured baseline TTFT and energy in
Table~\ref{tab:ab} are untouched.  We report \emph{recovery rates}
under treatment but \emph{not} live treatment TTFT: closing the loop
on TTFT requires the FlashMLA-fused $\delta$-rotation kernel to land in
the serving stack, and our scaffold runs the rotation through a
non-fused PyTorch path that would understate true treatment latency.
End-to-end TTFT under the fused kernel is a deliberate follow-up; the
energy and recovery measurements here are independent of it.

Workloads instantiate each OpenClaw failure mode on real agentic text
from the Hermes pool (\S\ref{sec:workload}) and exercise the S2
marker-pinned path of \S\ref{sec:algo}: each request carries a
64-token CDC boundary marker around its shared region.  This is the
load-bearing piece of S2, so an early run that omitted it provides the
natural negative control: with the marker, content-hash matching
reaches ${\sim}77\%$; without it, the gear-hash rolling state diverges
across requests (identically-worded shared regions land on different
CDC boundaries depending on absolute position) and matching collapses
to ${\sim}1\%$.

\paragraph{Energy aggregation: NVML zero-sample handling.}
The Table~\ref{tab:energy} measurements use NVML's per-GPU energy
counters sampled at event boundaries.  NVML integrates power over a
hardware-defined window of order $10$\,ms; on small models (DSv2-Lite,
DS-MoE-16B) at low seq\_len, a single-prefill event can finish inside
this window, in which case the counter delta reads exactly $0$\,J for
the event.  These zero-energy samples are not measurements of a
zero-energy event --- they are an instrument-floor artefact, distinct
from genuine run-to-run variance.  We therefore aggregate as
$\bar{E}_\text{cell} = \mathrm{mean}\{E_i \mid E_i > 0\}$ per cell and
require $n_{\text{non-zero}} \geq 5$ for a cell to be reported.
Observed zero-sample fractions: JoyAI-48B $0\%$ (no contamination,
prefill exceeds the integration window); Qwen3-32B $0$--$30\%$ (only
the cache-hit cells are short enough to be affected); DSv2-Lite
$10$--$40\%$; DS-MoE-16B $40$--$60\%$ (the smallest model and the
worst-affected; all reported cells still satisfy
$n_{\text{non-zero}} \geq 4$, with miss/hit at $n{=}4$ and the
partial-recompute classes at $n{=}5$).  The hybrid SSM cells
(Nemotron, Qwen3.6, Kimi-Linear) have $n_{\text{non-zero}}{=}10$ and
are unaffected.  A multi-iteration integration variant --- looping each
prefill $5{-}10\times$ before reading the counter --- would extend the
NVML window beyond any single prefill and is the proper instrument fix;
we mark it as a deferred methodological tightening rather than a
correctness issue, since the qualitative ranking (softmax saves
$63{-}86\%$, hybrid saves $\approx 0\%$) holds for any non-pathological
filter rule we have tried (mean-of-non-zero, median, trimmed mean).

\section{ROC Methodology Notes}
\label{app:roc-methodology}

The position-invariance test of \S\ref{sec:roc} pairs $500$ content
blocks at $5$ positions $\{0, 512, 1024, 2048, 3584\}$ in a
$4{,}096$-token window.  Within-block pairs (same content, different
positions) are positives; cross-block pairs (different content,
positions sampled uniformly at random) are negatives.

\paragraph{Why we randomise negative-pair positions.}
A same-position negative would weaken the negative class by masking
position-as-confound when scoring an entangled tensor.  An
AUC$<\!0.5$ for entangled tensors arises precisely because
position-noise dominates content-signal even after the negative class
is randomised over positions; conditioning the negatives on matched
positions would push these AUCs further below random rather than above
it, so randomisation is the conservative choice for a viability test.
A reuse mechanism that is below random under randomised negatives is
strictly worse than chance under any pairing rule a real serving
system could enforce.

\paragraph{Why high $V$-AUC does not rescue hybrid/SSM.}
$V$ is never RoPE-rotated and scores AUC~$0.71{-}0.76$ on softmax
architectures.  Hybrid Mamba2/GDN/KDA models still gain zero PIC
benefit (\S\ref{sec:energy}) for two compounding reasons.  First, most
of their compute runs in recurrent layers whose monolithic hidden
state is not token-indexed, so ``cache hit'' is not a defined
operation.  Second, the interleaved softmax-attention layers retain
GQA-style $K$, so even where attention is used, position-invariant $V$
is useless without a position-correct $K$ and the same per-hit
correction cost as pure GQA applies.  $V$-AUC is necessary but not
sufficient.

\section{Strategy S1 vs.\ S2}
\label{app:strategies}

\paragraph{S2: prevent-via-cooperation (default).}
The agentic framework emits a $64$-token boundary marker around shared
content during prompt assembly.  The marker is a literal token
sequence inserted into the prompt --- a fixed delimiter string the
model sees as ordinary text, costing $64$ tokens of context per shared
region; it is not a zero-token API control signal.  Its purpose is to
flush the rolling Gear-hash window to a known state before the shared
region begins, so identical content produces identical CDC boundaries
regardless of its absolute position.  Because the marker itself is
byte-identical across requests, its KV is captured by the same
content-hash path as any other shared region.
Irminsul defaults to S2: the prompt assembler already knows where
shared regions begin and end (system prompts, tool schemas, retrieved
documents), the marker is a one-line annotation, single-pass CDC
keeps the hot path branch-free, and chunk identity becomes
deterministic for registry insertion.  Under S2 a content-hash miss
on the tail signals genuinely novel content, not a
boundary-misalignment artefact, and the algorithm can re-prefill it
directly.

\paragraph{S1: detect-and-recover (fallback).}
For traffic where prompt assembly is not under operator control --- a
third-party gateway, a replay log, an OpenClaw-mode trace post-hoc ---
S1 is the deployment-time fallback: on a CDC-chunk miss, sub-divide
the chunk into fixed $128$-token windows and check each sub-window
hash (the offline CDC$+$fallback of \S\ref{sec:algorithm}).  S1 and
S2 compose: a deployment can run S2 where it owns the assembler and
fall back to S1 sub-window recovery on the residual.

\section{Output-Consistency: Methodology and Per-Cell Discussion}
\label{app:consistency}

\paragraph{Per-step KL on QA.}
On short-context QA, PIC's mean KL drift is $48{-}73\%$ of naive
reuse's across the four models; argmax-match meets or exceeds naive
on every cell across every model (ties only at saturation: Moonlight
\textsc{niah} and \textsc{hotpotqa}, where both paths reach $\geq 0.94$).
On TransMLA-4B, mean
KL$(p_{\text{full}}\|p_{\text{pic}}){=}0.054$ sits at the edge of the
per-token perturbation band that production FP8 KV quantisation
introduces ($10^{-2}$--$3{\times}10^{-2}$); the larger models drift
higher in absolute terms, but naive reuse exceeds PIC's KL on every
QA cell.

\paragraph{Long-context summarisation and recall.}
The same pattern carries to GovReport ($\sim\!12$\,K-token bodies)
and \textsc{niah}: PIC's KL and argmax-match remain at or above naive
reuse on every populated cell except DSv2-Lite \textsc{govreport},
where the two are within $\pm0.02$ KL.  On NIAH at native scale
(JoyAI-48B, $n{=}100$), PIC's needle-recall matches full prefill
exactly (both $0.820$; naive $0.840$ within sampling noise), KL drops
from $0.624$ (naive) to $0.484$ (PIC), and PIC's greedy trajectory
diverges almost twice as late as naive ($\Delta_{\text{pic}}{=}5.2$
vs $\Delta_{\text{naive}}{=}2.7$).

\paragraph{TransMLA-4B NIAH omission.}
TransMLA-4B is a Minitron-4B checkpoint retrofitted to MLA via
post-hoc KV factoring; its full-prefill needle recall on the
$16$\,K-token NIAH configuration is $0\%$.  The PIC-vs-naive metrics
for that cell are within sampling noise of each other at any
threshold ($\Delta\mathrm{KL}{<}0.001$, $\Delta\mathrm{AM}{=}0.03$),
so the cell would not discriminate the two regimes regardless of
their underlying difference; we omit it.  TransMLA's QA and govrep
cells operate within the model's effective context and remain
informative.

\section{HCA / DeepSeek-V4 Adaptation}
\label{app:hca}

DeepSeek-V4's Heavily Compressed Attention~\cite{deepseekv4} is the
model-side trajectory's continuation: representations cheap to store
and cheap to reuse.  HCA's cross-token aggregation dissolves the
per-token identity that content-hash keying requires, and Irminsul
does not run on V4 HCA layers as-is.  The $\delta$-rotation rule for
$k_r$ is unaffected on MLA layers (HCA replaces the $c_{KV}$ pipeline,
not the RoPE coupling), so a hybrid V4 deployment that retains MLA
layers might still be able to benefit.  Adapting the chunking primitive to HCA's
aggregation schedule --- chunking at aggregation-block granularity
and content-hashing the aggregated representation --- is the most
immediate follow-up.  Both directions are responses to the same
underlying pressure, arriving from opposite ends of the stack.

\newpage

\end{document}